# Teleportation of a two-particle entangled state via W class states

Zhuo-Liang Cao, Wei Song

Department of Physics, Anhui University, Hefei, 230039, P. R. of China

**Abstract**

A scheme for teleporting an unknown two-particle entangled state via W class states is proposed. In this scheme, the W class entangled states are considered as quantum channels. It is shown that by means of optimal discrimination between two nonorthogonal quantum states, probabilistic teleportation of the two-particle entangled state can be achieved.

## 1. Introduction

Quantum entanglement is one of the most striking features of quantum mechanics and has been used as an essential resource of quantum information processing such as quantum teleportation [1], quantum cryptography [2], quantum computation [3] and so on. Since the seminal work of Bennett et al [1], there have been extensive works in the field of quantum teleportation in theory and experiment [4-12]. Entangled states, which are called quantum channels, make it possible to send an unknown state a long distance.

Entanglement in three qubits is more complicated than that in two qubits. In Ref. [13], authors show that the entanglement of three qubits can be classified into GHZ states and W class state. The GHZ state cannot be transformed to the W class by the local operation and classical communication. Although many proposals have utilized the GHZ state in quantum teleportation, there few teleportation schemes using W class state in the paper.

Recently, Shi et al.[14] and J. Joo et al. [15] have proposed two different schemes to teleport the single particle state with W state. In Shi's proposal, the teleportation can be successfully realized with a certain probability if the receiver adopts an appropriate unitary-reduction strategy. In Joo's scheme, a sender performs positive operator valued measurement to realize the teleportation, but they only consider the question of how to teleport the single particle state, furthermore, the quantum channel are composed of W state. In this letter, we are interested in teleporting an unknown two-particle entangled state. The quantum channels are constructed by W class states. We show that the probabilistic teleportation of the original entangled state can be realized, by means of generalized quantum measurement.

## 2. Teleportation of an unknown two-particle entangled state

Firstly we set up a W class state to be used as quantum channels between Alice and Bob, which is in the following state

$$|\Psi\rangle_{123} = a|001\rangle_{123} + b|010\rangle_{123} + c|100\rangle_{123} \tag{1}$$

where $|a|^2 + |b|^2 + |c|^2 = 1$, $|a| > |b| > |c|$.

We suppose Alice has an entangled particle pair, which consists of particle 4 and 5. She wants to teleport the unknown state $|\Psi\rangle_{45}$ of the particle pair to Bob. The state $|\Psi\rangle_{45}$ may be expressed as

$$|\Psi\rangle_{45} = \alpha|00\rangle_{45} + \beta|11\rangle_{45} \tag{2}$$

where $|\alpha|^2 + |\beta|^2 = 1$. The particle 3 of the state $|\Psi\rangle_{123}$ and the particle pair (4, 5) belong to the sender Alice. Other two particles 1, 2 belong to receiver Bob. In order to realize the teleportation, a Bell measurement on particle 3 and 4 is made by Alice at the first step, which will project particles 1, 2 and 5 into the following states:

$$\langle\Phi^\pm|_{34}\Psi\rangle_{123} \otimes |\Psi\rangle_{45} = \pm\frac{a\beta}{\sqrt{2}}|001\rangle_{125} + \frac{b\alpha}{\sqrt{2}}|010\rangle_{125} + \frac{c\alpha}{\sqrt{2}}|100\rangle_{125} \tag{3}$$

$$\langle\Psi^\pm|_{34}\Psi\rangle_{123} \otimes |\Psi\rangle_{45} = \pm\frac{a\alpha}{\sqrt{2}}|000\rangle_{125} + \frac{b\beta}{\sqrt{2}}|011\rangle_{125} + \frac{c\beta}{\sqrt{2}}|101\rangle_{125} \tag{4}$$

where $|\Phi^\pm\rangle_{34} = \frac{1}{\sqrt{2}}(|00\rangle_{34} + |11\rangle_{34})$, $|\Psi^\pm\rangle_{34} = \frac{1}{\sqrt{2}}(|01\rangle_{34} + |10\rangle_{34})$. We find that if Bob operates a Von Neumann measurement on particle 1, the state of the particle 2 and 5 will be projected into the following

$$\langle 0|_1\langle\Phi^\pm|_{34}\Psi\rangle_{123} \otimes |\Psi\rangle_{45} = \pm\frac{a\beta}{\sqrt{2}}|01\rangle_{25} + \frac{b\alpha}{\sqrt{2}}|10\rangle_{25} \tag{5}$$

$$\langle 0|_1\langle\Psi^\pm|_{34}\Psi\rangle_{123} \otimes |\Psi\rangle_{45} = \pm\frac{a\alpha}{\sqrt{2}}|00\rangle_{25} + \frac{b\beta}{\sqrt{2}}|11\rangle_{25} \tag{6}$$

$$\langle 1|_1\langle\Phi^\pm|_{34}\Psi\rangle_{123} \otimes |\Psi\rangle_{45} = \frac{c\alpha}{\sqrt{2}}|00\rangle_{25} \tag{7}$$

$$\langle 1|_1\langle\Psi^\pm|_{34}\Psi\rangle_{123} \otimes |\Psi\rangle_{45} = \frac{c\beta}{\sqrt{2}}|01\rangle_{25} \tag{8}$$

From the above equations, we can see that if the result of the measurement on particle 1 is $|1\rangle_1$, the teleportation fails; if the result of the measurement on particle 1 is $|0\rangle_1$, the teleportation can be successful. But the equations (5) and (6) are still not the desired state. If Eq.(5) is obtained, Bob introduces an ancillary qubit (qubit A) in a state $|1\rangle_A$. Then the combined state is

$$|\Psi\rangle_{25} \otimes |1\rangle_A = \frac{1}{\sqrt{2}}\left(\pm a\beta|011\rangle_{25A} + b\alpha|101\rangle_{25A}\right) \qquad (9)$$

Bob performs a controlled-not operation(C-NOT) with particle 2 as the control bit and the ancillary particle A as the target bit, thus transforming the above state into the following one:

$$|\Psi\rangle_{25A} = \frac{1}{\sqrt{2}}\left(\pm a\beta|011\rangle_{25A} + b\alpha|100\rangle_{25A}\right) \qquad (10)$$

We note the state (10) can also be expressed as

$$|\Psi\rangle_{25A} = \frac{\sqrt{a^2+b^2}}{\sqrt{2}} \frac{1}{2}\bigl[\left(\pm \beta|01\rangle_{2A} + \alpha|10\rangle_{2A}\right)\left(x_1|1\rangle_5 + y_1|0\rangle_5\right)$$
$$\left(\pm \beta|01\rangle_{2A} - \alpha|10\rangle_{2A}\right)\left(x_1|1\rangle_5 - y_1|0\rangle_5\right)\bigr] \qquad (11)$$

where we assume $x_1 = \frac{a}{\sqrt{a^2+b^2}}, y_1 = \frac{b}{\sqrt{a^2+b^2}}$. At this stage Alice performs an optimal POVM [16] to conclusively distinguish between the two nonorthogonal quantum states $x_1|1\rangle_5 + y_1|0\rangle_5$ and $x_1|1\rangle_5 - y_1|0\rangle_5$. The respective positive operators that form an optimal POVM in this subspace are

$$A_1 = \frac{1}{2x_1^2}\begin{pmatrix} x_1^2 & x_1 y_1 \\ x_1 y_1 & y_1^2 \end{pmatrix},$$

$$A_2 = \frac{1}{2x_1^2}\begin{pmatrix} x_1^2 & -x_1 y_1 \\ -x_1 y_1 & y_1^2 \end{pmatrix},$$

$$A_3 = \begin{pmatrix} 0 & 0 \\ 0 & 1-\frac{y_1^2}{x_1^2} \end{pmatrix}. \qquad (12)$$

The probability of the optimal state discrimination from such an generalized measurement is $2y_1^2$, if Alice confirms that after the POVM measurement of the state of particle 5 is $x_1|1\rangle_5 + y_1|0\rangle_5$ $\left(x_1|1\rangle_5 - y_1|0\rangle_5\right)$, she can informs Bob the result of her measurement through classical channel. Then Bob confirms that the quantum state composed of particles 2 and A is $\pm\beta|01\rangle_{2A} + \alpha|10\rangle_{2A}$ $(\pm\beta|01\rangle_{2A} - \alpha|10\rangle_{2A})$

For the outcome $\pm\beta|01\rangle_{2A} + \alpha|10\rangle_{2A}$, Bob performs a unitary operation $(\pm|0\rangle\langle 1| + |1\rangle\langle 0|)_2 \otimes I_A$ on the state, then transforming it into $\alpha|00\rangle_{2A} + \beta|11\rangle_{2A}$, which is the desired state corresponding to faithful teleportation. When the outcome is $\pm\beta|01\rangle_{2A} - \alpha|10\rangle_{2A}$, Bob performs a unitary operation $(\pm|0\rangle\langle 1| - |1\rangle\langle 0|)_2 \otimes I_A$ on the state, the teleportation still can be realized successfully.

If Eq.(6) is obtained, Bob introduces an ancillary qubit in a state $|0\rangle_A$. Then the combined state is

$$|\Psi\rangle_{25} \otimes |0\rangle_A = \frac{1}{\sqrt{2}}(\pm aa|000\rangle_{25A} + b\beta|110\rangle_{25A}) \tag{13}$$

Bob still performs a controlled-not operation(C-NOT) with particle 2 as the control bit and the ancillary particle A as the target bit, the state will be transformed into the following one

$$|\Psi\rangle_{25A} = \frac{1}{\sqrt{2}}(\pm aa|000\rangle_{25A} + b\beta|111\rangle_{25A}) \tag{14}$$

It can be expressed as

$$|\Psi\rangle_{25A} = \frac{\sqrt{a^2+b^2}}{\sqrt{2}} \frac{1}{2}[(\pm\alpha|00\rangle_{2A} + \beta|11\rangle_{2A})(x_1|0\rangle_5 + y_1|1\rangle_5)$$
$$(\pm\alpha|00\rangle_{2A} - \beta|11\rangle_{2A})(x_1|0\rangle_5 - y_1|1\rangle_5)] \tag{15}$$

where the values of $x_1$ and $y_1$ are the same as above. The optimal POVM operation for discriminating between the two quantum states $x_1|0\rangle_5 + y_1|1\rangle_5$ and $x_1|0\rangle_5 - y_1|1\rangle_5$ becomes

$$B_1 = \frac{1}{2x_1^2}\begin{pmatrix} y_1^2 & x_1 y_1 \\ x_1 y_1 & x_1^2 \end{pmatrix},$$

$$B_2 = \frac{1}{2x_1^2}\begin{pmatrix} y_1^2 & -x_1 y_1 \\ -x_1 y_1 & x_1^2 \end{pmatrix},$$

$$B_3 = \begin{pmatrix} 1 - \frac{y_1^2}{x_1^2} & 0 \\ 0 & 0 \end{pmatrix} \quad (16)$$

The optimal probability of the state identification is $2y_1^2$. With similar analysis, the teleportation can be successfully realized. The unitary transformation corresponding to the state of particles 2 and A are given in Table 1

**Table 1. Unitary transformations corresponding to the state of particles 2 and A**

| States of particles 2 and A | Bob's Unitary transformation |
|---|---|
| $\beta\|01\rangle_{2A} + \alpha\|10\rangle_{2A}$ | $(\|0\rangle\langle 1\| + \|1\rangle\langle 0\|)_2 \otimes I_A$ |
| $-\beta\|01\rangle_{2A} + \alpha\|10\rangle_{2A}$ | $(-\|0\rangle\langle 1\| + \|1\rangle\langle 0\|)_2 \otimes I_A$ |
| $\beta\|01\rangle_{2A} - \alpha\|10\rangle_{2A}$ | $(\|0\rangle\langle 1\| - \|1\rangle\langle 0\|)_2 \otimes I_A$ |
| $-\beta\|01\rangle_{2A} - \alpha\|10\rangle_{2A}$ | $(-\|0\rangle\langle 1\| - \|1\rangle\langle 0\|)_2 \otimes I_A$ |
| $\alpha\|00\rangle_{2A} + \beta\|11\rangle_{2A}$ | $(\|0\rangle\langle 0\| + \|1\rangle\langle 1\|)_2 \otimes I_A$ |
| $-\alpha\|00\rangle_{2A} + \beta\|11\rangle_{2A}$ | $(-\|0\rangle\langle 0\| + \|1\rangle\langle 1\|)_2 \otimes I_A$ |
| $\alpha\|00\rangle_{2A} - \beta\|11\rangle_{2A}$ | $(\|0\rangle\langle 0\| - \|1\rangle\langle 1\|)_2 \otimes I_A$ |
| $-\alpha\|00\rangle_{2A} - \beta\|11\rangle_{2A}$ | $(-\|0\rangle\langle 0\| - \|1\rangle\langle 1\|)_2 \otimes I_A$ |

If a successful teleportation occurs, the unknown two-particle entangled state can be reproduced on Bob's side with fidelity 1. The total probability of the successful teleportation is $P = \frac{a^2 + b^2}{2} \times \frac{1}{4} \times 2 \times 2y_1^2 \times 4 = 2|b|^2$, we can see when the parameters $|c|$ is small enough the probability will approach 1.

### 3. Conclusion

In summary, we have proposed a simple scheme of teleporting an unknown two-particle entangled state via W class states. In this scheme, the quantum channel is constructed by W class states, which make it more general than W state. We show that if Alice performs an optimal POVM to distinguish between the two nonorthogonal quantum states, then she inform Bob the result of her measurement through classical channel, a probability of teleportation can be realized successfully, and the fidelity in our scheme can reaches 1, which is higher than that in Ref.[12]. Now, the preparation of W state have been discussed in Ref.[17,18], so our scheme may be realized in experiment.

**Acknowledgement**

This work is supported by the Natural Science Foundation of Anhui Province under Grant No: 03042401 and key program of the Education Department of Anhui Province.